\documentclass[prb,twocolumn,showpacs,preprintnumbers,amsmath,amssymb]{revtex4}
\usepackage{dcolumn}
\usepackage{bm}
\usepackage{graphicx}

\begin{document}
\hyphenation{Hof-stad-ter}
\preprint{Klinkhammer et al.}
\title{Magnetoresistance of  antidot lattices with grain boundaries} 
\author{S. Klinkhammer}\author{Hengyi Xu}\author{T. Heinzel} \email{thomas.heinzel@uni-duesseldorf.de}
\affiliation{Heinrich-Heine-Universit\"at, Universit\"atsstr. 1,
40225 D\"usseldorf, Germany} \author{U. Gennser}\author{G.
Faini}\author{C. Ulysse} \author{A. Cavanna}\affiliation{CNRS-LPN, Route de Nozay,
91960 Marcoussis, France} \date{\today}

\begin{abstract}
The magnetotransport properties of antidot lattices containing
artificially designed grain boundaries have been measured. We find
that the grain boundaries broaden the commensurability resonances and displace them anisotropically. These phenomena are unexpectedly weak but differ
characteristically from isotropic, Gaussian disorder in the antidot
positions. The observations are interpreted in terms of
semiclassical trajectories which tend to localize along the grain
boundaries within certain magnetic field intervals. Furthermore, our results indicate how the transport
through superlattices generated by self-organizing templates may get
influenced by grain boundaries.
\end{abstract}

\pacs{73.23.-b, 73.63.-b}

\maketitle

\section{\label{sec:1}INTRODUCTION}

Superlattices imposed on two-dimensional electron gases (2DEGs) by
lateral patterning have revealed many fascinating effects
over the past two decades. One version of such superlattices are
antidot lattices,  where  the potential  maxima lie above the Fermi
energy. \cite{Ensslin1990,Weiss1991,Lorke1991} The
longitudinal magnetoresistance shows resonances at magnetic fields
where the cyclotron orbit is commensurate with the lattice
\cite{Weiss1991},  an effect which originates from the classically
chaotic character of such systems, as demonstrated by simulations
based on the Kubo formalism. \cite{Kubo1957,Fleischmann1992}  This
model has also been used to explain the nonlinear Hall resistance of
antidot lattices, \cite{Fleischmann1994} which can even develop a
pronounced negative Hall resistance at small magnetic fields.
\cite{Weiss1991} Many quantum effects in lateral superlattices have
been reported as well. For example, $B$-periodic oscillations,
superimposed on the commensurability oscillations, have been
explained  by quantization along closed orbits
\cite{Weiss1993,Richter1995} in square lattices. In hexagonal
lattices, a transition from Altshuler-Aronov-Spivak oscillations at
small magnetic fields to Aharonov-Bohm oscillations at larger fields
has been observed. \cite{Nihey1995,Ueki2004,Iye2004} Moreover, the
long-standing prediction of the Hofstadter butterfly,
\cite{Hofstadter1976} the fractal energy spectrum of a periodic
potential in strong magnetic fields, has been experimentally
confirmed in weak periodic superlattices.
\cite{Geisler2004,Albrecht2001,Schlosser1996} Antidots have
furthermore been used to demonstrate the quasi-particle character of
composite fermions in a very intuitive way. \cite{Kang1993}

These results have generated the quest to drive lateral
superlattices deeper into the quantum regime which requires smaller
lattice constants. With established techniques reaching their limits
around periods of 100 nm, \cite{Geisler2004,Schlosser1996}
self-organizational schemes offer themselves as an alternative.
Potential techniques include masks from porous alumina
\cite{Liang2002} or from diblock copolymers \cite{Thurn2000}.
Recently, Melinte et al. demonstrated that a diblock copolymer
system in the spherical phase can be used to generate a
two-dimensional lateral superlattice in a Ga[Al]As heterostructure
with lattice constants below 40 nm. \cite{Melinte2004} Diblock
copolymer systems in the hexagonal columnar phase have already been
used to pattern, e.g., ferromagnet/normal metal hybrid systems
\cite{Bal2002} as well as semiconductor surfaces. \cite{Black2001}
However, in many self-organizing systems, domain formation and the
corresponding disorder at the grain boundaries (GBs) poses a
limitation to the periodicity of the patterns.
\cite{Thurn2000,Melinte2004} Based on this background, the question
how GBs in two-dimensional superlattices influence the transport
properties is a relevant one which has not yet been addressed,
to the best of our knowledge. Beyond that, the study of such
mesoscopic GBs is of fundamental relevance. First of all, GBs in natural
polycrystalline conductors usually trap impurities, which may be charged and dominate
the influence of the GBs on the conduction
electrons.\cite{Grovenor1985,Nakamichi1996} The artificially
designed, mesoscopic GBs in lateral superlattices, on the other
hand, are free of such impurities and therefore allow us to study the
purely geometrical impact of a GB on the transport
properties. Second, the electronic
wave packets in antidot lattices can be highly localized on the scale of the lattice constant and thus within the GB, a situation which is not possible in conventional conductors.

Here, we report an investigation of artificial but
realistic GBs defined in a hexagonal antidot lattice. The GBs have
been generated by Monte-Carlo simulations with appropriate boundary
conditions, followed by a pattern transfer into the 2DEG via conventional lithographic
techniques. We observe a surprisingly weak, but nevertheless
characteristic, anisotropic damping and shift of commensurability resonances
in the magnetoresistance. These
experimental results are supported by numerical simulations within
the Kubo formalism. The calculations reveal that the resistance anisotropy has
their origin in a small, magnetic field dependent fraction of
electron trajectories that tend to localize along the GBs over large magnetic field intervals.

The paper is organized as follows. In Section II, we present the
sample preparation and the experimental setup. Section III contains
the experimental results and their interpretation. Section IV gives
a summary and conclusions.

\section{\label{sec:2}SAMPLE PREPARATION AND EXPERIMENTAL SETUP}

Conventional modulation-doped $\,\mathrm{GaAs/ Al_{0.2}Ga_{0.8}As}$
heterojunctions with a two-dimensional electron gas (2DEG) $100\,\mathrm{nm}$
below the surface were grown by molecular beam epitaxy. The pristine
2DEG has an electron density of $n=3.2 \times
10^{15}\,\mathrm{m^{-2}}$ and a mobility of
$\mu=68\,\mathrm{m^2 /Vs}$, corresponding to an elastic mean free path of
$l=6.4\,\mathrm{\mu{m}}$ and a Drude scattering time of
$\tau_{D}=26\,\mathrm{ps}$ at liquid helium temperatures. Since all
components of the (anisotropic) resistivity tensor must be
accessible, we defined a square ($20\,\mathrm{\mu{m}} \times
20\,\mathrm{\mu{m}}$) van der Pauw-geometry with 8 contacts by
electron beam lithography and subsequent ion beam etching with
low-energy ($250\,\mathrm{eV}$) $Ar^+$ ions.\cite{Lee1997} The etch
depth was $20\,\mathrm{nm}$ which is sufficient to deplete the 2DEG.
For Ohmic contacts to the 2DEG we alloyed Ni/AuGe pads in the
heterostructure.

\begin{figure}[htb]
\centering
\includegraphics{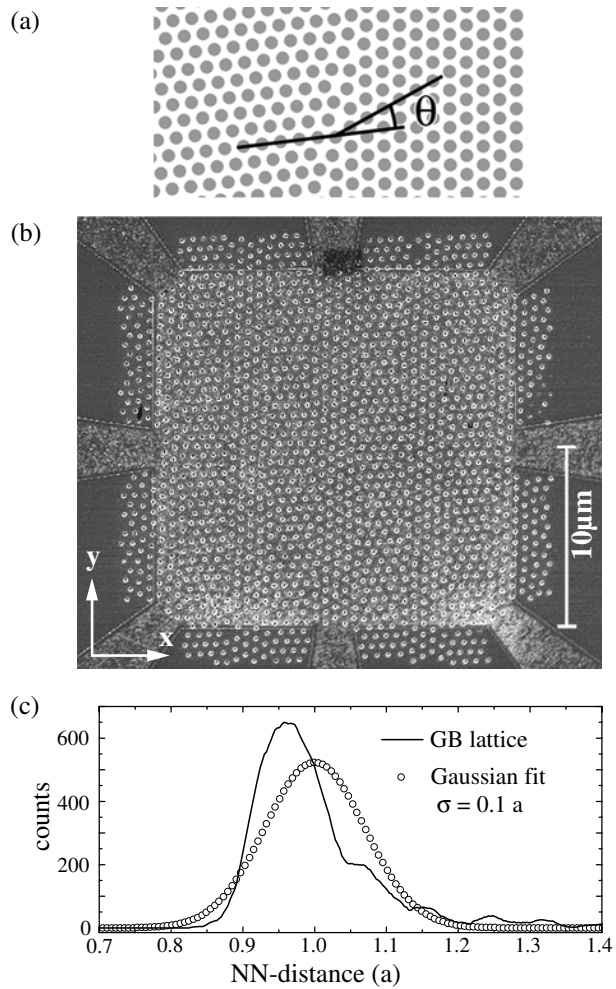}
\caption{(a) A section of the hexagonal lattice generated by the
Monte-Carlo simulation containing a $\Sigma 7$ grain boundary, characterized by
the angle $ \theta = 21.8\,\mathrm{^\circ}$. (b) Top view of the patterned sample prior to the deposition of the top gate, as seen in a scanning electron microscope. A stripe of the lattice around the GB along the y direction is replicated across the van der Pauw square in the center of the mesa that contains the 2DEG (light gray). (c) Histogram of the
nearest-neighbor distances of the designed GB lattice, measured in units of the average lattice constant $a$. For this purpose, the replication of the GB lattice has been extended to $10^4$ antidots. A Gaussian
fit gives a  mean value of 1 with amplitude and $\sigma=0.10$ in units of $a$ as fit parameters. }\label{Antidots_GB_Fig1}
\end{figure}

The 2D lattice containing the GB was numerically generated by using
a standard Metropolis Monte-Carlo simulation algorithm with periodic
boundary conditions and a repulsive, two-dimensional
Yukawa-Potential which is cut off after three lattice constants.
\cite{Honerkamp2002} Without any initial conditions (i.e. \emph{seeds}), this
would generate a perfectly hexagonal lattice. In order to enforce
the formation of a GB, we defined two seeds, each consisting of 19
sites arranged in perfectly hexagonal geometry, separated by 20
lattice constants and rotated to each other by an angle of
$\theta=21.8\,\mathrm{^\circ}$, see Fig. \ref{Antidots_GB_Fig1}(a), which corresponds to the $\Sigma 7$ coincidence
site lattice within one of the standard
classification schemes for GBs \cite{WolfMerkle1992,Bonnet1980},
meaning that one out of 7 lattice sites coincide in an overlay of
the two grains. It belongs
to the set of energetically favorably GBs according to various
models,\cite{Palumbo1992} and occurs frequently in real systems see,
e.g. Puglisi et al.\cite{Puglisi2007}, Fig. 3(a) or Black et
al.\cite{Black2007}, Fig. 1(c). We believe, however, that the
experimental results reported below are rather insensitive to the
kind of coincidence site lattice, and we note that we are unaware of
a scheme for classifying GBs according to the microscopic details
of the local geometry at the boundary, which would probably be most
appropriate for our experiments.

The simulated lattices with GBs were transferred to the pattern
generator of an electron beam writer. The GBs were oriented parallel to the mesa edge in y-direction. The superlattices were
patterned by electron beam lithography and ion beam etching as
described above. The average lattice constant was
$a=600\,\mathrm{nm}$ with a lithographic antidot diameter of
$d_{lith}=200\,\mathrm{nm}$. We opted for patterning antidot
lattices with a large ratio $d_{eff}/a\approx 0.7$, where $d_{eff}$
is the effective electronic antidot diameter, in order to amplify
the effects due to the disorder at the GBs as compared to lattices with smaller $d_{eff}/a$ ratios, at the expense of the number of observable commensurability oscillations in similarly
patterned arrays.\cite{Meckler2005} Finally, on two of the samples, a Ti/Au
($20\,\mathrm{nm}/200\,\mathrm{nm}$) top gate was evaporated.

The magnetoresistance was measured in a helium gas flow cryostat with a
variable temperature insert and a superconducting magnet with a
maximum field of $B=8\,\mathrm{T}$ which is applied perpendicular to
the plane of the 2DEG. For temperatures at $100\,\mathrm{mK}$, a $\rm{^3He/^4He}$ dilution
refrigerator was used. All measurements were performed in a
four-probe setup using standard lock-in techniques with a current of
$100\,\mathrm{nA}$ at frequencies of $13.7\,\mathrm{Hz}$.

We did not observe any GB effect on antidot lattices
containing just one GB along the y-direction as indicated in Fig. \ref{Antidots_GB_Fig1}(a) (not shown). Therefore, the simulated GB was periodically replicated without disturbing the
regular hexagonal lattice parts outside the grain boundary area, see
Fig. \ref{Antidots_GB_Fig1}(b). The average distance between two GBs
was $d_{GB} \approx 2.5\,\mathrm{\mu{m}}$. This is about $40\%$ of
the mean free path of the unpatterned 2DEG and well above the
effective mean free path in the antidot lattice, estimated from
the resistivity at zero magnetic field to $\approx 280\,\mathrm{nm}$ . Therefore, we can exclude
the possibility of commensurability effects due to the
one-dimensional GB superlattice. At the same time, $d_{GB}$ is
larger than  the spatial extension of a GB, such that its character
is retained.
Fig. \ref{Antidots_GB_Fig1}(c) shows the distribution of nearest neighbor (NN) distances in the GB lattice. The NN distances show marked deviations from Gaussian disorder. Compared to a Gaussian NN distribution with the same average distance $a$ and the same single standard deviation of $0.1 a$, a long tail of large distances is observed, while the majority of the NN distances is slightly reduced, resulting in a histogram peak at a NN distance of $\approx 0.96 a$, in combination with a suppression of very small NN distances. This indicates that within the domain wall, a few antidots with large NN distances are present wich compress the surrounding lattice locally. Very small NN distances below $\approx 0.9 a$, however, are suppressed, due to the repulsive potential used in the Monte Carlo simulation. As a result, the maximum of the measured NN distribution is larger than the corresponding Gaussian distribution, and the peak appears significantly sharper around its maximum.

Four samples of this geometry and of nominally identical values $d_{lith}/a$ have
been measured. They all show qualitatively very similar behavior. Here,
we show data from the two samples where the effects were most pronounced.
In addition, two single domain hexagonal lattices in the same sample
have been measured for control purposes.

\section{\label{sec:3}EXPERIMENTAL RESULTS AND INTERPRETATION}

\begin{figure}[htb]
\centering
\includegraphics{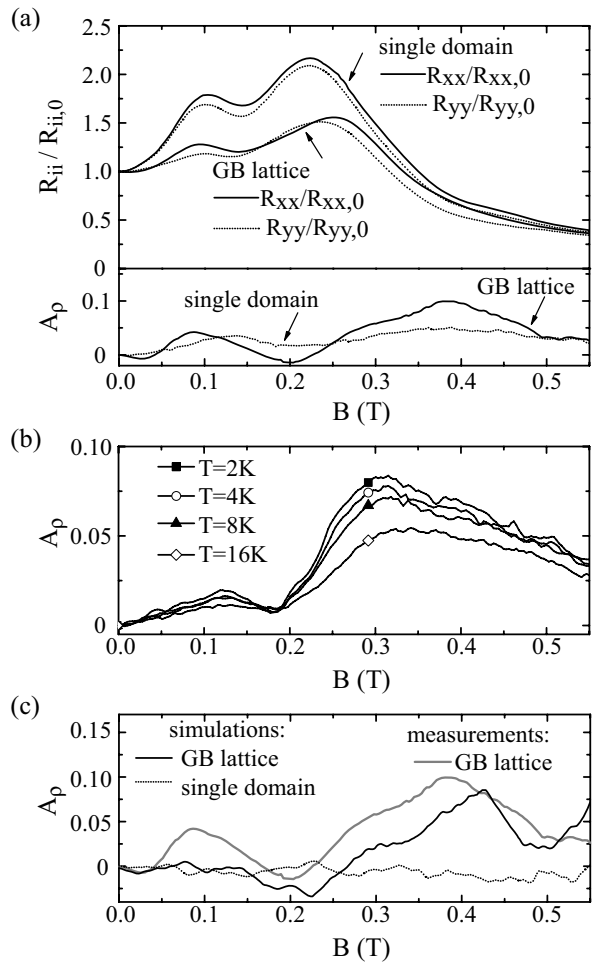}
\caption{(a) Upper part: longitudinal magnetoresistances of a
single domain hexagonal antidot array and a GB superlattice. Both
samples were ungated. Lower part:  the corresponding  resistance
anisotropy $A_{\rho}$ as defined by Eq. \ref{Antidots_GB_Fig1}. (b) Temperature
dependence of $A_{\rho}$ for a second sample containing a GB lattice.
(c) Comparison of the experimentally determined $A_{\rho}$ of the GB lattice shown in (a) with the simulations.}
\label{Antidots_GB_Fig2}
\end{figure}

In Fig. \ref{Antidots_GB_Fig2}(a), the longitudinal magnetoresistances
$R_{xx}$ and $R_{yy}$ for a single domain hexagonal antidot lattice and
for the lattice containing the GBs (as reproduced in Fig. \ref{Antidots_GB_Fig1}) are shown. For better comparison all measurements were normalized to their zero-field value.

The measurements show the characteristic commensurability oscillations of antidot arrays. Within the simple
commensurability picture,  the cyclotron radius of the
electrons  roughly matches the
orbital radius around 1 and 7 antidots at $B\approx 230\,\mathrm{mT}$ and $\approx 100\,\mathrm{mT}$, respectively.\cite{Weiss1994} The resonance around 3 antidots is weakly pronounced and hidden in the wing towards lower magnetic fields of the resonance around one antidot. The suppression of this resonance is also observed in our simulations for antidots with  large $d/a$ ratios and is a simple geometrical effect. We also
observe a weak resonance at $B\approx 450\,\mathrm{mT}$ which corresponds to
electrons caught in between three adjacent antidots, as possible in
arrays with large $d/a$ ratios.\cite{Meckler2005} Moreover, the characteristic
overall resistance decrease as a function of increasing  $B$
is observed, which ends roughly where the
cyclotron radius equals the spatial extension of the electron gas
between two adjacent antidots. This was used to
estimate the effective antidot diameter to $d_{eff}\approx
460\,\mathrm{nm}$ up to $d_{eff}\approx 520\,\mathrm{nm}$ depending on the sample and the top gate voltage.

The observation of the commensurability oscillations demonstrates their resilience with respect to GBs.  However, compared  with the single domain lattice, the commensurability peaks in the GB lattice are significantly damped and somewhat broadened. The peak position in the GB lattice that corresponds to the cyclotron motion around a single antidot is shifted to larger magnetic fields, while the positions of the other two resonances are approximately constant. This peak shift is about 5\% for $R_{yy}$ and 10\% for $R_{xx}$, and cannot be explained with deviations of  the lattice constants from the designed values, since the accuracy of the electron beam writer allows for 1\% deviation at most. Rather, we interpret this displacement as a consequence of the NN distribution. The sharp peak of NN distances at a reduced value corresponds to a locally reduced lattice constant, which is reflected in the position of  the resonance around a single antidot while the resonance around 7 antidots is less sensitive to this effect since it averages over more NN distances.

The single domain lattice shows
isotropic behavior to a good approximation, whereas in the lattice containing the GBs,
the transport is anisotropic. All three resonances are slightly suppressed in the y-direction, i.e., along the grain boundaries, as compared to the x-direction (perpendicular to the GBs). Especially the the commensurability resonance at $B \approx 240\,\mathrm{mT}$ is shifted to smaller magnetic fields in the y direction. Furthermore, $R_{yy}$ is significantly lower than $R_{xx}$ at and above the resonance around a single antidot. We emphasize that this phenomenology is observed in all our GB lattices. This anisotropy somewhat resembles that one reported by Tsukagoshi et al. \cite{Tsukagoshi1995}, who measured rectangular antidot lattices with uni-directional, Gaussian disorder and observed a damping of
the commensurability oscillations as well as a shift to smaller magnetic fields for the current perpendicular to the disorder direction. In analogy to this behavior, one is tempted to conclude that in the GB lattice, the disorder perpendicular to the GBs is greater than parallel to them. However, we could not think of a way to check this, e.g. by  looking at the x- and y-components of the NN distributions, since our single domain stripes have different orientations with respect to each other and many antidots at the grain boundaries cannot be unambiguously attributed to one domain.

We define the resistivity anisotropy $A_{\rho}$ as

\begin{equation}
\label{DWAntidotsEq1}A_{\rho}\equiv \frac{\rho_{xx} -
\rho_{yy}}{\rho_{xx} + \rho_{yy}}
\end{equation}

In the van der Pauw geometry used, $A_{\rho}$ is equivalent to a
correspondingly defined resistance
anisotropy \cite{Vries1995,Pauw1958} and nicely illustrates the anisotropy effects described above. In the lower part of Fig. \ref{Antidots_GB_Fig2}(a) and in Fig. \ref{Antidots_GB_Fig2}(b), $A_{\rho}(B)$ for our
measurements is reproduced. While $A_{\rho}$ for the single domain lattice shows only a weak magnetic field dependence and stays always smaller than 5\%, it shows peaks at the commensurability resonances and a broad shoulder extending from $\approx 250\,\mathrm{mT}$ to $\approx 500\,\mathrm{mT}$, with a maximum up to 10 \% in the GB lattice. Moreover, $A_{\rho}$ has a minimum close to $B= 200\,\mathrm{mT}$, where the absolute values can even get negative. Fig.
\ref{Antidots_GB_Fig2}(b) shows that the temperature dependence of $A_{\rho}$ between $2\,\mathrm{K}$ and
$16\,\mathrm{K}$ is quite weak, suggesting a classical origin. Additional measurements at
$T=0.1\,\mathrm{K}$ in a different cooldown (not shown) revealed no significant change of the effects as compared to $2\,\mathrm{K}$.

We proceed by interpreting our experimental results with the help of
semiclassical simulations based on the Kubo
formalism.\cite{Kubo1957} The magnetoconductivity tensor was calculated in the lattice geometries as
patterned in the heterostructure. The antidots were modeled as
hard-wall circles with diameters similar to $d_{eff}$, namely $460\,\mathrm{nm}$. For each
value of the magnetic field, we computed the trajectories of
$10^5$ electrons at the experimentally determined Fermi level of
$E_F = 10\,\mathrm{meV}$ in this potential landscape. The step
width for the magnetic field was $\Delta{B} = 2\,\mathrm{mT}$. The
electrons start at random directions of the Fermi velocity and at
random positions within an area of $10\times 20$ antidots. This
large start area proved to be necessary to capture the transport
properties of the whole GB within acceptable computation times. The electron trajectories are computed
to a length of $75\,\mathrm{\mu{m}}$ from which we calculate the
velocity correlation function $C_{ij}(t,B)\equiv \langle
v_i(t,B)v_j(0,B)\rangle$, $(i,j = x,y)$ for every value of B.\cite{Richter2000} Here,
the brackets denote averaging over all trajectories. The components $ \sigma_{ij}(B)$
of the magnetoconductivity tensor for a degenerate two-dimensional
electron gas follow from

\begin{equation}
\label{DWAntidotsEq2} \sigma_{ij}(B) = \frac{m^* e^2}{\pi {\hbar}^2}
\int\limits_0^\infty \langle v_i(t,B)v_j(0,B)\rangle e^{-t/\tau_{D}}
\mbox{d}t
\end{equation}

Here, $m^*=0.067m_e$ represents the effective
electron mass in GaAs. In order to
compare our experimental results with the simulations, the
conductivity tensor is converted into a resistivity tensor.

The simulations agree reasonably well with  the main structure of the measured $A_{\rho}(B)$, as can be seen in Fig. \ref{Antidots_GB_Fig2}(c). The peak around $0.1\,\mathrm{T}$, the minimum close to $0.2\,\mathrm{T}$ followed by a peak around  $0.4\,\mathrm{T}$  are reproduced, with an amplitude close to the measured one. Since the simulations are carried out at zero temperature, this agreement supports our observation that thermal smearing is relatively unimportant. The simulated resistivity anisotropy for a single domain lattice, both without disorder as well as with isotropic disorder with a standard deviation of $\sigma =0.1 a $ (not shown), does not show any structure, in accordance with experimental results obtained by us as well as by others.\cite{Takahara1995} However, the shift of the commensurability resonance around one antidot towards larger magnetic fields as compared to the single domain lattice is not observed in the simulations. We speculate that in addition to the qualitative argument given above, a  more realistic confinement potential is necessary to generate this effect.

\begin{figure}[htb]
\centering
\includegraphics[scale=0.5]{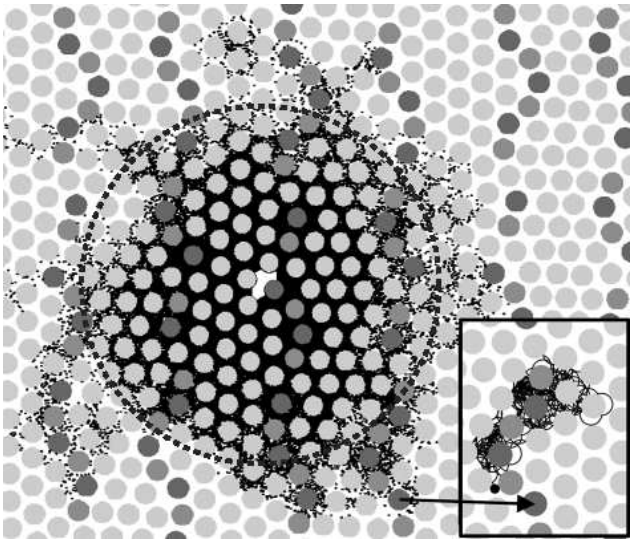}
\caption{Main figure: the diffusion cloud formed by 300 electron trajectories that started from one unit cell  (the white area at the center of the array) at a magnetic field of $B=400\,\mathrm{mT}$. For better visibility, only every $50^{th}$ trajectory point, corresponding to a trajectory length of $250\,\mathrm{nm}$, is shown. Inset: zoom of one trajectory bound to a GB. Note its
localization at antidots with a coordination number different from 6
(the gray tone of an antidot represents its coordination
number. Dark gray: $z=7$, light gray: $z=6$, and medium gray: $z=5$). The dashed circle indicates the edge of an isotropic diffusion cloud.}\label{Antidots_GB_Fig3}
\end{figure}

We cannot resort to the conventional method of Poincar$\rm{\acute{e}}$
sections \cite{Fleischmann1992} for further analysis, since our lattice containing the GBs is not periodic.
We therefore follow a more qualitative approach and inspect the
longitudinal velocity correlation functions of individual
trajectories, given by $\tilde{C}_{ii}(B)\equiv \int
v_i(t,B)v_i(0,B)e^{-t/\tau_{D}}dt$. For different magnetic fields, we
identified  trajectories that contribute most to $A_{\rho}$. Such
trajectories were found by inspection of  the differences between $\tilde{C}_{yy}(B)$  and $\tilde{C}_{xx}(B)$  in the magnetic field intervals where the maximum of  $A_{\rho}$ occurs, namely around $B= 400\,\mathrm{mT}$.  The result is exemplified in Fig. \ref{Antidots_GB_Fig3}, where we plot 300 trajectories starting from one unit cell. The GBs are visualized by representing the coordination numbers $z$ of the antidots in gray scale. Coordination numbers of 5 and 7 are present in the GBs. The diffusion cloud has an anisotropic shape, in contrast to the corresponding cloud for a single domain lattice (not shown). Rather, the GBs form weak diffusion barriers, and characteristic branches of trajectories are visible that extend along the GBs.

It is found that most of the trajectories in the branches are localized at a GB for a long time. The inset in  Fig. \ref{Antidots_GB_Fig3} shows one example of such a trajectory. It can be seen that the electrons show a tendency to circle around antidots with coordination numbers $z$ of 5 or 7 for some time, before they move on to a neighboring antidot, which often has a $z\neq 6$ as well. For the simulations with the parameters of our samples, this behaviour is typical for magnetic fields above $0.3\,\mathrm{T}$. At smaller magnetic fields, $A_{\rho}$ is significantly smaller and even can become negative around $B\approx 0.2\,\mathrm{T}$, which indicates that the electrons diffuse better across the GB than along it. However, we could not relate this behavior unambiguously to a particular type of trajectory in the corresponding diffusion clouds, probably because the effect is too weak.

\section{\label{sec:4}SUMMARY AND CONCLUSION}
Longitudinal resistances parallel and perpendicular to artificially
generated grain boundaries in a hexagonal antidot lattice imposed on
a 2DEG in a semiconductor heterostructure have been measured. We observe that in comparison to a single domain lattice, the
commensurability resonances broaden and develop a pronounced anisotropy. This suggests that
the disorder across the grain boundaries is stronger than along
them. The
effect of the grain boundaries on the electron trajectories has been visualized by numerical
studies and consists of a slight localization of the electrons along the grain
boundaries for most, but not all, magnetic fields. In a small magnetic field interval in between the resonances around one antidot and seven antidots, both the experiments and the simulations suggest that an enhanced diffusion across the grain boudaries is possible as well.

We simulated various GB types and could not identify
characteristic microconfigurations of antidots at the boundary.
Therefore, we believe that the effects reported are qualitatively
independent of the GB type. Future work could address several  issues  in further detail. Our choice of the ratio antidot diameter/lattice constant was based on a heuristic argument in combination
with our experimental constraints. It is unclear how the effects
change with this parameter. Also, the influence of the modulation
strength is an open question. In particular, it will be worth
looking at weakly modulated systems, which are more
likely to be obtained by self-organizational schemes.
\cite{Melinte2004} The exact shape of the antidot potential,
however, seems of minor relevance, since our samples have large
lateral depletion lengths and with it a soft potential, and
nevertheless the measured resistance anisotropies agree well with the simulations carried out on hard-wall arrays.

In superlattice structures prepared by self-organization where GBs
might matter, their type, density and orientation is per se at best
marginally under control, and in samples containing many grains, the
observed broadening of the resonances will be isotropic. However,
the disorder generated by the GBs is non-Gaussian and locally anisotropic and thus different from random disorder. Our results indicate that even a high density of grain boundaries does not destroy the specific transport properties of the superlattice, at least in the classical regime.

The authors would like to thank C. Likos for support and valuable
discussions regarding the Monte Carlo simulations. This work was
financially supported by the CNRS-LPN and the
Heinrich-Heine-Universit\"at D\"usseldorf.
\newpage

\end{document}